\documentclass[12pt]{article}
\usepackage{epsfig}
\usepackage{graphicx}
\usepackage{a4}
\usepackage{latexsym}
\usepackage{cite}

\usepackage{pslatex}
\usepackage[latin1]{inputenc}
\usepackage[T1]{fontenc}

\newcommand{\eqmel}{\raisebox{-0.07cm}{$\:\stackrel{{\rm M}}{=}\:$} }

\newcommand{\beq}{\begin{equation}}
\newcommand{\eeq}{\end{equation}}
\newcommand{\bea}{\begin{eqnarray}}
\newcommand{\eea}{\end{eqnarray}}
\newcommand{\nn}{\nonumber}
\newcommand{\ra}{\rightarrow}
\newcommand{\MSb}{$\overline{\mbox{MS}}$}
\newcommand{\as}{\alpha_{\rm s}}
\newcommand{\ar}{a_{\rm s}}
\newcommand{\ep}{\epsilon}

\newcommand{\ec}{\gamma_e}

\def\ca{{C^{}_A}}
\def\cas{{C^{\,2}_A}}
\def\cat{{C^{\,3}_A}}
\def\cf{{C^{}_F}}
\def\cfs{{C^{\, 2}_F}}
\def\cft{{C^{\, 3}_F}}
\def\cff{{C^{\, 4}_F}}
\def\nf{{n^{}_{\! f}}}

\def\tf{{T^{}_{\! f}}}

\def\x1{{(1 \! - \! x)}}

\def\Ms{{M^{\:\! 2}}}
\def\xs{{x^{\, 2}}}
\def\frct#1#2{\raisebox{-0.5mm}{\mbox{\large{$\:\!\frac{#1}{#2}\:\!$}}}}

\begin{document}
\setlength{\parskip}{0.2cm}
\setlength{\baselineskip}{0.53cm}

\begin{titlepage}

\noindent
IPhT-T14/084, LTH 1013 \hspace*{\fill} July 2014\\
DESY 14-106, LPN14-084 \\
\vspace{1.5cm}
\begin{center}
\Large
{\bf Leading large-$x$ logarithms of the quark-gluon contributions\\[1mm]
to inclusive Higgs-boson and lepton-pair production}\\
\vspace{2cm}
\large
N.A. Lo Presti$^{\, a}$, A.A. Almasy$^{\, b\,\ast}$ and A. Vogt$^{\, c}$\\
\normalsize
\vspace{1cm}
{\it $^a$Institut de Physique Th\'eorique, CEA-Saclay\\[0.3mm]
F-91191, Gif-sur-Yvette cedex, France}\\
\vspace{0.4cm}
{\it $^b$ Deutsches Elektronensynchrotron DESY \\[0.3mm]
Platanenallee 6, D--15738 Zeuthen, Germany}\\
\vspace{0.4cm}
{\it $^c$Department of Mathematical Sciences, University of Liverpool \\[0.3mm]
Liverpool L69 3BX, United Kingdom}\\[2cm]
\vfill
\large
{\bf Abstract}
\vspace{-0.2cm}
\end{center}
We present all-order expressions for the leading double-logarithmic threshold 
contributions to the quark-gluon coefficient functions for inclusive 
Higgs-boson production in the heavy top-quark limit and for Drell-Yan 
lepton-pair production. These results have been derived using the structure 
of the unfactorized cross sections in dimensional regularization and the 
large-$x$ resummation of the gluon-quark and quark-gluon splitting functions. 
 The resummed coefficient functions, which are identical up to
colour factor replacements, are similar to their counterparts in deep-inelastic
scattering but slightly more complicated.
\vspace{1.0cm}

\noindent
$^\ast$ {\small Address until 31 August 2013} 
\end{titlepage}


\noindent
The discovery of a particle with a mass of about 125 GeV \cite{LHC2012} and 
properties consistent with those of the standard-model Higgs boson 
\cite{LHCstatus} at the LHC has led to increased interest in precision 
predictions for Higgs production and decay.
The main channel for the total production cross section is gluon-gluon fusion 
via a top quark loop, known at all $M_{\rm H}/M_{\rm top}$ to next-to-leading 
order (NLO) of perturbative QCD \cite{HiggsNLO,Harlander:2005rq}. 
The convergence of the perturbation series is particularly slow in this case,
hence calculations are required at, and beyond, the next-to-next-to-leading 
order (NNLO).

These calculations can be carried out, at a sufficient accuracy
\cite{HiggsMtop}, for an effective $Hgg$ interaction in the heavy-top limit 
\cite{HGGeff},
\beq
\label{L-Hgg} 
   {\cal L}_{\,\rm eff} \;\: = \;\: - \frct{1}{4} \: C_H \: H \,
   G^{\,a}_{\!\mu\nu} G^{\,a,\,\mu\nu} \:\: ,
\eeq
where $G^{\,a}_{\!\mu\nu}$ denotes the gluon field strength tensor. The 
prefactor $C_H$ includes all QCD corrections to the top quark loop; it is of 
first order in the strong coupling constant $\as$ and fully known up to N$^3$LO
($\,\as^{\,4\,}$) \cite{Chetyrkin:1997un}, see also Refs.~\cite{CHgg05}.
The NNLO contributions to the total cross sections were computed in this 
effective theory in Refs.~%
\cite{Harlander:2002wh,Anastasiou:2002yz,Ravindran:2003um};
a high-accuracy threshold resummation and a first approximation for N$^3$LO 
corrections were subsequently obtained in Refs.~\cite{Catani:2003zt,MV1}.

Recently a major step has been taken towards deriving the complete N$^3$LO 
corrections: the calculation of the soft-gluon and virtual contributions at 
this order \cite{Anastasiou:2014vaa}. This result directly leads to a further
improvement in the threshold limit
\cite{Ahmed:2014cla,Bonvini:2014joa,Catani:2014uta}
by fixing the remaining parameter required for a full N$^3$LO + 
next-to-next-to-next-to-leading logarithmic (N$^3$LL) accuracy \cite{MVV7} of 
the soft-gluon exponentiation.
The same soft$\,+\,$virtual N$^3$LO and resummation accuracy has also been 
reached for Drell-Yan lepton-pair production 
  $pp \ra \ell^+ \ell^- +$ {\it anything}, 
calculated at NNLO in Refs.~\cite{Hamberg:1991np,Harlander:2001is}, due to its 
close similarity with inclusive Higgs-boson production 
\cite{Ahmed:2014cla,Catani:2014uta}.

Generally fixed- or all-order results for logarithmically enhanced 
endpoint contributions, e.g., in the large-$x$ or threshold limit, can provide
checks of elaborate Feynman-diagram calculations and estimates of corrections 
that cannot (yet) be calculated directly. 
Quite a few studies of the threshold limit have addressed the
dominant channels in Higgs and lepton-pair production, i.e., gluon-gluon 
fusion and quark-antiquark annihilation, respectively.
Here we present first all-order results for the sub-dominant quark-gluon 
contributions to both processes. In particular, we derive the leading large-$x$
logarithms of the coefficient functions $c_{P,\rm qg}^{}$ for $P = \rm H$ and 
$P = \rm DY$.

\vspace{3mm}
Our derivation starts from the unfactorized partonic cross sections 
$\widehat{W}_{P,j\ell}^{}$ in 
\beq
\label{Sunfact}
  \sigma_P^{} \;\;=\;\; 
     \widetilde{\sigma}_{0,P}^{} \: \widehat{W}_{P,j\ell}^{\,} 
     \:\otimes\: \widehat{f}_{\!j}^{} \:\otimes\: \widehat{f}_\ell^{}
  \;\;=\;\; \widetilde{\sigma}_{0,P}^{} \; \widetilde{c}_{P,ik}^{} 
     \:\otimes\: Z_{ij} \:\otimes\: Z_{k\ell} 
     \:\otimes\: \widehat{f}_{\!j}^{} \:\otimes\: \widehat{f}_\ell^{} \:\: ,
\eeq
which lead to the mass-factorized expressions
\beq
\label{Sfact}
  \sigma_P^{} \;\;=\;\; \sigma_{0,P}^{} \: c_{P,ik}^{} 
     \:\otimes\: f_i^{} \:\otimes\: f_k^{} \:\: .
\eeq
Here $\otimes$ abbreviates the Mellin convolutions, and summations over the 
light quarks and antiquarks and gluons are understood. All charge factors 
have been suppressed; see, e.g., Appendix A of Ref.~\cite{Hamberg:1991np} for 
the Drell-Yan process.
We use dimensional regularization with $D = 4 - 2\:\!\ep$; a tilde marks 
the $D$-dimensional counterparts of quantities which are finite for $\ep = 0$. 
In particular, the coefficient functions in Eq.~(\ref{Sunfact}) can be written 
as 
\beq
\label{Dcoeff}
   \widetilde{c}_{P,ik}^{}(x,\Ms) \;\;=\;\; \sum_{n=0}\; \sum_{\ell = 0}\: 
   \ar^{\,n}\, \ep^{\,\ell}\, c_{P,ik}^{(n,\ell)}(x)
 \quad \mbox{ with } \quad \ar \equiv \frct{\as(\Ms)}{4\pi}
\eeq
for the choice $\mu_r^{} = \mu_f^{} = M$ of the renormalization and 
mass-factorization scales, with $M = M_H$ or $M = M_{\ell^+\ell^-}$, which can
by made without loss of information. 
All factorized expressions refer to the \MSb\ scheme; the additional terms 
defining its difference to MS are suppressed in Eq.~(\ref{Dcoeff}) and below.
The coefficient functions $c_{P,ik}^{}$ in Eq.~(\ref{Sfact}) are obtained from 
the above by setting $\ep = 0$.

The scale dependence of the factorized parton distributions $f_i^{}$ in 
Eq.~(\ref{Sfact}) is governed by the splitting functions $P_{ik}^{}$, which 
are related to the transition functions $Z_{ik}$ in Eq.~(\ref{Sunfact}) by
\beq
\label{PofZ}
  P_{ik}^{} \;\: \equiv \;\: - \,\gamma_{ik}^{}
  \;\:=\;\: \frac{d\:\! Z_{ij} }{d\ln \Ms } \:\otimes\: [Z^{\,-1}]_{jk}
  \;\:=\;\: \beta_D^{}(\ar) \;\frac{d\:\! Z_{ij} }{d \ar} \:\otimes\:
  [Z^{\,-1}]_{jk}
\:\: ,
\eeq
where $\beta_D^{}(\ar) = -\,\ep\, \ar - \beta_{0\,}^{} \ar^{\,2} - \ldots$ with 
$\beta_0 = \frac{11}{3}\,\ca - \frac{2}{3}\,\nf$ is the $D$-dimensional beta 
function.  Eq.~(\ref{PofZ}) can be solved for $Z$ order by order in $\as$.
 
The prefactors $\widetilde{\sigma}_{0,P}^{}$ in Eq.~(\ref{Sunfact}) are defined 
such that the lowest-order contributions to the \mbox{$D$-dimensional} 
coefficient functions in Eq.~(\ref{Dcoeff}) are normalized and independent of 
$\ep$, i.e., given~by 
\beq
\label{cHDY0}
  c_{\rm H,gg}^{\,(0,\ell)}(x) \;\;=\;\;
  c_{\rm DY,q\bar{q}}^{\,(0,\ell)}(x) \;\;=\;\; 
  \delta\x1 \: \delta_{\,0\ell} \:\: .
\eeq
%
%
We further specify our notation for the coefficient functions and splitting 
functions by recalling the leading-logarithmic large-$x$ contributions to the 
NLO quark-gluon coefficient functions:
\bea
\label{cHqg1} 
  c_{\:\!\rm H,qg}^{\,(1)\,\rm LL}(x)  &\!=\!& 
    2\:\! P_{\rm gq}^{\,(0)}(x) \: \ln \x1 
    \;\;=\;\; 
    4\, \cf ( \,2\:\!x^{\,-1} - 2 + x\, ) \: \ln \x1 \:\: ,
\\[1.5mm]
\label{cDYqg1} 
  c_{\rm DY,qg}^{\,(1)\,\rm LL}(x)  &\!=\!& 
    2\:\! P_{\rm qg}^{\,(0)}(x) \, \ln \x1
    \;\;=\;\; 
    \, 4\, \tf \: (\,1 - 2\:\!x + 2\:\!\xs\,) \; \ln \x1 
\eea
with $\cf = \frac{4}{3}$, $\tf = \frac{1}{2}$ and $\ca = 3$ for QCD.
Note that our convention in Eq.~(\ref{cHqg1}) differs from the quantities 
$\Delta_{\,ik}$ in Refs.~\cite{Anastasiou:2002yz,Ravindran:2003um} by a factor 
of $x^{\,-1}$. On the other hand, our normalization in Eq.~(\ref{cDYqg1}) is 
the same as in  Ref.~\cite{Hamberg:1991np}. The corresponding NNLO corrections
read
\bea
\label{cHqg2} 
  c_{\:\!\rm H,qg}^{\,(2)\,\rm LL}(x) \; &\!=\!&
    \frct{1}{3} \left( 13\,\cf \,+\, 35\,\ca \right) 
    \, P_{\rm gq}^{\,(0)}(x) \: \ln^3 \x1 \:\: , 
\\[1.5mm]
\label{cDYqg2} 
  c_{\rm DY,qg}^{\,(2)\,\rm LL}(x) \; &\!=\!&
    \frct{1}{3} \left( 35\,\cf \,+\, 13\,\ca \right)
    \, P_{\rm qg}^{\,(0)}(x) \: \ln^3 \x1 \:\: .
\eea

\vspace{3mm}
It is convenient to turn the convolutions above to products by Mellin 
transforming all quantities, 
\beq
\label{Mtrf}
 f(N) \;=\; \int_0^1 \! dx \,
 \left(\, x^{\,N-1} \{ - 1 \} \right) \: f(x)_{\{+\}}
 \:\: ,
\eeq
where the parts in curly brackets refer to the case of $(1-x)^{-1}$
$+$-distributions. 
Here we mainly consider the leading powers of $\x1$ in the threshold limit,
in particular $\x1^0$ corresponding to $N^{\,-1}$ in the large-$N$ limit for 
the quark-gluon quantities addressed in this letter.
Keeping only the leading -- and subleading, if $\ln^{\,k} N$ is replaced by
$\ln^{\,k} N + k\,\ec \ln^{\,k-1} \!N$ -- contributions, the relations between 
the corresponding expressions in $x$-space and Mellin-$N$ space read
\beq
\label{Logtrf}
  \frac{\ln^{\,n} \!\x1}{\x1_+} \;\;\eqmel\;\;
  \frac{(-1)^{n+1}}{n+1} \, \ln^{\,n+1} \!N \:+ \:\ldots
\:\: , \quad
  \ln^{\,n} \!\x1  \;\;\eqmel\;\;
  \frac{(-1)^n}{N} \, \ln^{\,n} \! N  \: + \:\ldots \;\; .
\eeq
Here and below $\eqmel$ denotes equality under the Mellin transformation 
(\ref{Mtrf}).

The diagonal splitting function are not logarithmically enhanced at higher
orders for the $N^{\,0}$ contributions \cite{Korchemsky:1989si} 
(nor at $N^{\,-1}$, see Refs.~\cite{MVV34,DMS05}$\,$). 
Hence only their leading-order contributions are relevant here (and at NLL), 
with
\beq
\label{Pii0}
  P_{\rm qq}^{\,(0)\,\rm LL}(N) \;\: = \;\: -\, 4\,\cf \ln N 
\;\; , \quad
  P_{\rm gg}^{\,(0)\,\rm LL}(N) \;\: = \;\: -\, 4\,\ca \ln N
\:\: .
\eeq
The corresponding off-diagonal contributions can be readily read off from
Eqs.~(\ref{cHqg1}) and (\ref{cDYqg1}),
\beq
\label{Pij0}
   P_{\rm qg}^{\,(0)\,\rm LL}(N) \;\; = \;\; 2\:\! \tf \: N^{\,-1} 
\;\; , \quad
   P_{\rm gq}^{\,(0)\,\rm LL}(N) \;\; = \;\; 2\, \cf \: N^{\,-1}
\:\: .
\eeq
These functions do exhibit a double-logarithmic higher-order 
enhancement, derived in Ref.~\cite{AV2010},
\bea
\label{Pqgres}
  P_{\rm qg}^{\:\rm LL}(N,\ar) &\! = \! & 
  \ar \, P_{\rm qg}^{\,(0)\,\rm LL}(N) \: {\cal B}_{\,0}(-\tilde{a}_{\rm s})
\:\: , \\[1.5mm]
  P_{\rm gq}^{\:\rm LL}(N,\ar) &\! = \! &
  \ar \, P_{\rm gq}^{\,(0)\,\rm LL}(N) \; {\cal B}_{\,0}(\tilde{a}_{\rm s})
\label{Pgqres}
\eea
in terms of the function
\beq
\label{B0}
  {\cal B}_{\,0}(x) 
  \;\: = \;\: \sum_{n=0}^\infty \,\frac{B_n}{(n!)^2} \; x^{\,n}
  \;\: = \;\: 1 \,-\: {x \over 2} \; - \;
         \sum_{n=1}^\infty \,\frac{(-1)^n}{[(2n)!]^{\,2}}\; |B_{2n}|\, x^{\,2n}
\;\; ,
\eeq
where $B_n$ are the Bernoulli numbers in the standard normalization of 
Ref.~\cite{AbrSteg}, and
\beq
\label{astilde}
   \tilde{a}_{\rm s} \;\: \equiv \;\:
   4\, \ar \, (\cf\! - \:\!\!\ca)  \ln^{\,2} N \:\: .
\eeq
For the corresponding NLL and NNLL resummations of the splitting functions see 
Refs.~\cite{ASV10,PijNN14}.

\vspace{3mm}
We are now prepared to return to the unfactorized cross sections in 
Eq.~(\ref{Sunfact}). For brevity the following steps are written out only for
Higgs-boson production. 
We have checked that the corresponding relations for the Drell-Yan case can
be obtained, as expected from Eqs.~(\ref{cHqg1}) -- (\ref{cDYqg2}) and
(\ref{Pii0}) -- (\ref{astilde}), by interchanging gluon and (anti-)$\,$quark
indices and colour factor replacements.

For the resummation of the quark-gluon coefficient function 
$c_{\rm H, qg}^{} = c_{\rm H, \bar{q}g}^{}$ 
we need to consider
\beq
\label{WHqghat}
  \widehat{W}_{\rm H, qg}^{\,} \;\; = \;\:
  {\cal O}(N^{\,-1}) \;\;=\;\; 
  \widetilde{c}_{\rm H, qg}^{} \: Z_{\rm qq} \: Z_{\rm gg}
  \,+\, 
  \widetilde{c}_{\rm H, gg}^{} \: Z_{\rm gq} \: Z_{\rm gg}
  \:+\: {\cal O}(N^{\,-3})
\eeq
and 
\beq
\label{WHgghat}
  \widehat{W}_{\rm H, gg}^{\,} \;\: = \;\: 
  \; {\cal O}(N^{\,0}) \; \;\;=\;\;
  \widetilde{c}_{\rm H, gg}^{} \:\ Z_{\rm gg} \: Z_{\rm gg}
  \:+\: {\cal O}(N^{\,-2})
\hspace*{2.5cm}
\eeq
%
which provides $\widetilde{c}_{\rm H, gg}^{}$ for the right-hand-side of 
Eq.~(\ref{WHqghat}).
Other coefficient functions such as $\widetilde{c}_{\rm H, q\bar{q}}^{}$ are 
not relevant for the leading logarithms in Eq.~(\ref{WHqghat}) even at higher 
orders in $N^{\,-1}$.
 
At the leading (and next-to-leading) power in $N^{\,-1}$ the $\ar^{\,n}$
contributions to the diagonal and off-diagonal transition functions are 
given by \cite{AV2010}
\bea
\label{ZiiLL}
  Z_{ii}^{\,(n)\,\rm LL} &\! = \!&
  {1 \over n!}\: \ep^{-n} \left( \gamma^{\,(0)}_{\,ii} \right)^n
\:\: , \\[1mm] 
  Z_{ik}^{\,(n)\, \rm LL} &\! = \!&
  {1 \over n!} \;\sum_{m=0}^{n-1} \ep^{-n+m} \;\sum_{\ell=0}^{n-m-1}\;
    {(m+\ell)! \over \ell!}
    \left( \gamma^{\,(0)}_{\,ii} \right)^{n-m-\ell-1} \gamma^{\,(m)}_{\,ik}
    \left( \gamma^{\,(0)}_{\,kk} \right)^\ell
\:\: .
\label{ZikLL}
\eea
Here additional sign factors have been avoided by using the anomalous 
dimensions $\gamma$ defined in Eq.~(\ref{PofZ}).
The $D$-dimensional coefficient function $\widetilde{c}_{H, \rm gg}^{}$
can be determined from Eq.~(\ref{WHgghat}) with
\beq
\label{WHggD} 
  \widehat{W}_{\rm H, gg}^{\,\rm LL} \;\; = \;\;
  \exp \left( \ar \widehat{W}_{H,\rm gg}^{\,(1)\,\rm LL} \right)
\eeq
and
\beq
\label{WH1ggD} 
  \widehat{W}_{\rm H, gg}^{\,(1)\,\rm LL} 
  \;\; = \;\;
  4\, \cf \:{1 \over \ep^2}\: \left( \exp\, (2\:\! \ep \ln N) - 1 \right)
  \;\;\eqmel\;\; 
  - 4\: \cf {1 \over \ep}\: \x1_+^{-1-2\:\! \ep} + \,\mbox{virtual}
\eeq
at order $N^{\,0}$. The difference of Eq.~(\ref{WH1ggD}) to the corresponding 
structure function in deep-inelastic scattering (DIS) is the replacement 
$\ep\ra 2\,\ep$ in the exponentials due to the different phase space. An 
extension of Eqs.~(\ref{ZiiLL}) -- (\ref{WH1ggD}) to higher logarithmic
accuracy is no problem, but not required here.

\vspace{3mm}
The right-hand-side of Eq.~(\ref{WHqghat}) is thus known at LL accuracy at
all powers of $\as$ and $\ep$ except for the quark-gluon coefficient function. 
Hence an all-order result for $\widehat{W}_{H,\rm qg}^{}$ on the left-hand-side
corresponding to Eqs.~(\ref{WHggD}) and (\ref{WH1ggD}) leads to a LL 
resummation of $c_{\rm H, qg}^{}$; determining this result is the crucial step 
of our calculations.

Taking into account $\x1^{-k\,\ep}$ factors due to real and virtual 
corrections, cf.~the discussion of the phase-space master integrals in 
Ref.~\cite{Anastasiou:2002yz}, the general form of the $\ar^{\,n}$ 
contribution to $\widehat{W}_{H,\rm qg}^{}$ is
\bea
\label{WHqgD} 
  \widehat{W}_{\rm H, qg}^{\,(n)} &\! = \!&
  \frac{1}{\ep^{\,2n-1}} \: \sum_{\ell=2}^{2n} \: \x1^{-\ell\,\ep}
  \left ( \bar{A}_{\rm H, qg}^{(n,\ell)} 
        \:+\: \ep\, \bar{B}_{\rm H, qg}^{\,(n,\ell)} \:+\: \ldots \right)
  \; + \; {\cal O} \left( \x1^{\,1-k\,\ep} \right)
\nn \\ &\! \eqmel \!&
  \frac{1}{N\,\ep^{\,2n-1}} \: \sum_{\ell=2}^{2n} \; e^{\,\ell\,\ep\, \ln N}
  \,\left ( A_{\rm H, qg}^{(n,\ell)}
        \:+\: \ep\, B_{\rm H, qg}^{\,(n,\ell)} \:+\: \ldots \right)
  \; + \; {\cal O} \left( N^{\,-2} \,e^{\,k\,\ep\,\ln N} \right) 
\:\: .
\eea
The parameters $A_{\rm H, qg}^{(n,\ell)}$ combine to the coefficients of the 
LL contributions $\ar^{\,n}\; \ep^{\,-2\,n+m}\; \ln^{\,m-1}\! N$ in 
Eqs.~(\ref{WHqghat}), which, of course, vanish for $1 \,\leq\, m \,\leq\, n-1$ 
due to Eqs.~(\ref{ZiiLL}) and (\ref{ZikLL}).
Correspondingly, the quantities $B_{\rm H, qg}^{(n,\ell)}$ determine the NLL
contributions at all powers of $\as$ and $\ep$.

The presence of $2n-1$ terms in the sums (\ref{WHqgD}) represents a crucial 
difference to $\widehat{W}_{\rm H, gg}^{\,(n)}$ in the $N^{\,0}$ soft-gluon
limit, where only the $n$ even values of $\ell$ occur \cite{MV1}, and inclusive
DIS and semi-inclusive $e^+e^-$ annihilation (SIA), where the corresponding 
sums run from $\ell = 1$ to $\ell = n$ \cite{ASV10,Radcor11}. 
In those cases, an N$^n$LO calculation leads to a N$^n$LL resummation with a 
large number of relations to spare. Here, instead, all $2n-1$ terms with 
negative powers 
of $\ep$ are required to fix the LL coefficients $A_{\rm H, qg}^{(n,\ell)}$,
i.e., the terms to $\ep^{\,-2}$ fixed by lower-order contributions together
with the $\ep^{\,-1}$ term provided by the splitting-function resummation 
(\ref{Pgqres}).
Consequently, due to the extra factor of $\ep$, the NLL coefficients 
$B_{\rm H, qg}^{(n,\ell)}$ in Eq.~(\ref{WHqgD}) cannot be determined without
additional information.

\pagebreak

\begin{figure}[t]
\centerline{\epsfig{file=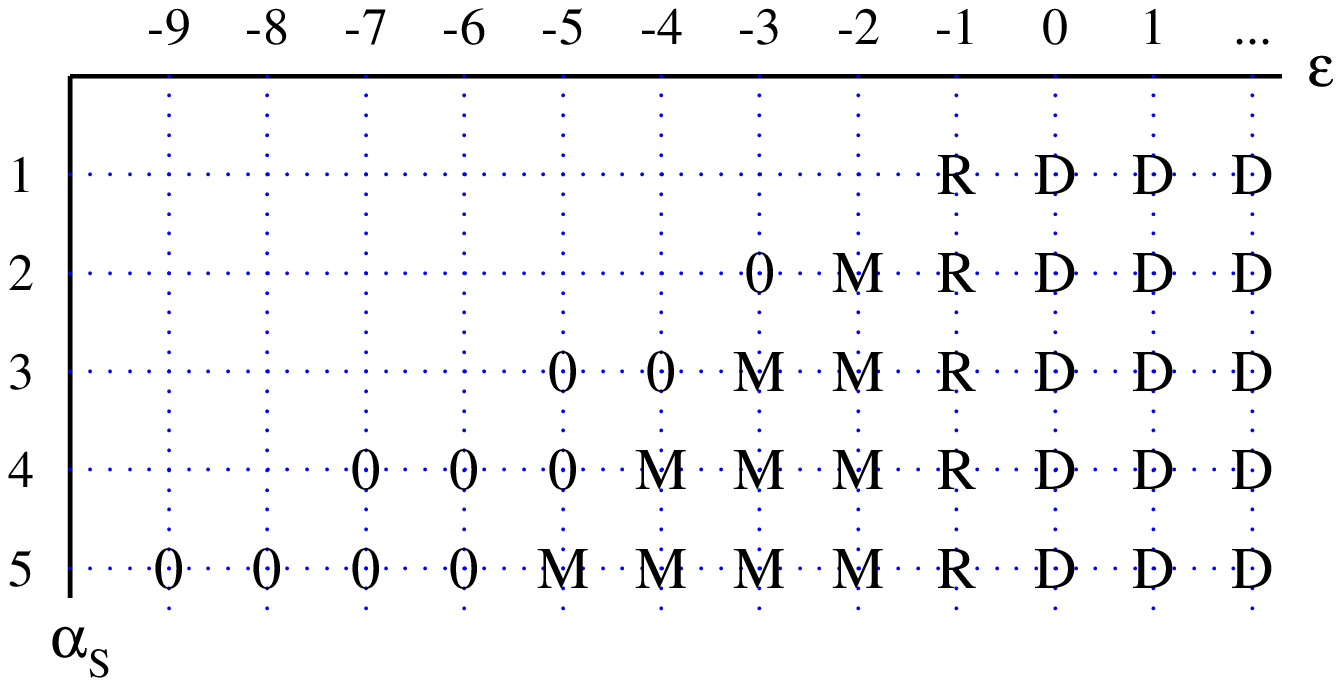,width=12.0cm,angle=0}}
\vspace{-3mm}
\caption{ \label{fig1}
The origin of the LL coefficients of $\ar^{\,n}\,\ep^{\,k}$ in 
Eqs.~(\ref{WHqghat}) and (\ref{WHqgD}) for $n \leq 5$.
`0' indicates double-pole combinations of $n$ and $k$ which are present in 
the latter but not the former equation.  
Entries marked by `M' are fixed by lower-order quantities through the mass 
factorization formula. 
The $\ep^{\,-1}$ terms (`R') are required at each order to determine the 
$2\:\!n-1$ coefficients $A_{\rm H, qg}^{(n,\ell)}$, they involve the splitting 
functions provided by fixed-order calculations at $n \leq 3$ and the 
resummations (\ref{Pqgres}) and~(\ref{Pgqres}). Finally entries marked by
`D' are determined, at each order, from the above coefficients via 
Eq.~(\ref{WHqgD}).
Checks of this procedure are provided by the $\ar^{\,2}\,\ep^{\,0}$ terms of 
Refs.~\cite
{Harlander:2002wh,Anastasiou:2002yz,Ravindran:2003um,Hamberg:1991np,%
Harlander:2001is}, see Eqs.~(\ref{cHqg2}) and (\ref{cDYqg2}), and the 
$\ar^{\,2}\,\ep^{\,1}$ contributions to Higgs production calculated 
in Ref.~\cite{Hoschele:2012xc}.
}
\vspace{-2mm}
\end{figure}

We have determined the coefficients $A_{\rm H, qg}^{(n,\ell)}$ in 
Eq.~(\ref{WHqgD}) to a sufficiently high order in $\as$ and~find
\bea
\label{Alist}
  A_{\rm H, qg}^{(n,2)} &\! = \!& 2\,\cf \;\frac{(-1)^{\,n}}{(n-1)!}
  \; (4\,\ca)^{\,n-1} 
\;\: , \nn \\[1mm]
  A_{\rm H, qg}^{(n,3)} &\! = \!& 2\,\cf \;\frac{(-1)^{\,n}}{(n-2)!}
  \; 2\,(\cf - \ca) \, (4\,\ca)^{\,n-2}
\;\: , \nn \\[1mm] 
  & \ldots &
\nn \\ 
  A_{\rm H, qg}^{(n,2\:\!n)} &\! = \!&  2\,\cf \;\frac{-1}{n!} \; 
  \sum_{k=0}^{n-1} \; (4\,\ca)^{\,k} \, (4\,\cf)^{\,n-1-k}
\:\; ,
\eea
which can be cast in a closed, if not very transparent, form in terms 
of binomial coefficients:
\beq
\label{Aclosed}
  A_{\rm H, qg}^{(n,\ell)} \; = \;
    \frac{4^{\,n}}{2\,n!} \:
   \sum_{m=1}^{\lfloor \ell/2 \rfloor} (-1)^{n+m+1}
      \left( {\begin{array}{c} n \\ \!\! \ell-m \!\! \\ \end{array}} \right)
     \: \sum_{k=0}^{m-1} \:
     \left( {\begin{array}{c} \!\! \rho+k \!\! \\ k \\ \end{array}} \right)
     (\cf - \ca)^{\rho} \, C_F^{\,k+1} \, C_A^{\,n-k-\rho-1} 
\eeq
with $\rho = \ell-2m$ and $\lfloor a \rfloor$ the largest integer not greater 
than $a$. 
The simplicity of especially the special cases (\ref{Alist}) provides 
some additional insurance against calculational errors. 
It is interesting to note that not only $A_{\rm H, qg}^{(n,3)}$, but all 
odd-$\ell$ coefficients vanish for $C_F = C_A$. 

\vspace{3mm}
With these results the LL mass-factorization of $\widehat{W}_{\rm H, qg}^{}$ 
can be performed order by order; it leads to a table of coefficients 
%
which has been given to $n=12$ in Ref.~\cite{NALP2012}. Finally this table 
can be used to find and verify the all-order resummation formula for the 
quark-gluon coefficient functions, 
\beq
\label{cHqgLL}
   c_{\rm H, qg}^{\,\rm LL_{\,}}(N,\ar)  \; = \;
   {1 \over 2\,N \ln N} \:\frct{\cf}{\cf - \ca} \left\{ 
   \exp \,(8\,\ca\, \ar \ln^{\,2} N)\, {\cal B}_{\,0} (\tilde{a}_{\rm s})
   - \exp \,((2\,\ca + 6\,\cf)\, \ar \ln^{\,2} N) 
   \right\}
\:\: ,
\eeq
which involves the same ingredients as its counterpart for DIS \cite{AV2010}
but is slightly more complicated. The corresponding coefficient function for
the Drell-Yan process can be obtained from (\ref{cHqgLL}) by 
$\cf \ra \tf$ in the numerator of the prefactor and $\ca \leftrightarrow \cf$
everywhere else, including the argument of the function ${\cal B}_{\,0}$.
Expansion of Eq.~(\ref{cHqgLL}) and Mellin inversion yields the explicit
third- and fourth-order predictions
\bea
  c_{\rm H, qg}^{(3)\,\rm LL}(x,\ar)  &\! = \!&
  \ln^5 \x1 \, \Big( 
    18\, \cft \,+\, \frct{100}{3}\,\cfs\,\ca \,+\, \frct{230}{3}\, \cf\,\cas 
  \Big) 
\:\: , \\[1.5mm]
  c_{\rm H, qg}^{(4)\,\rm LL}(x,\ar)  &\! = \!&
  \ln^7 \x1 \, \Big( 
    \frct{3646}{135}\, \cff \,+\, \frct{2834}{45}\,\cft\,\ca \,+\, 
    \frct{3166}{135}\, \cfs\,\cas \,+\, \frct{24434}{135}\,\cf\,\cat 
  \Big)
\:\:  \quad
\eea
and their obvious analogues for lepton-pair production. 

To summarize, we have derived the leading-logarithmic large-$x$ resummation of 
the quark-gluon coefficient functions for inclusive Higgs-boson and lepton-pair
production; our main results are Eq.~(\ref{cHqgLL}) and its closely related
counterpart for the Drell-Yan process.
Our calculations have been confined to the leading term in the expansion in 
powers of $\x1$; yet we definitely expect the structure with $P_{ik}
^{\,(0)}(x)$ in Eq.~(\ref{cHqg1}) -- (\ref{cDYqg2}) to occur at all orders.
An extension of our results to the next-to-leading double logarithms, 
$\as^{\,n} \ln^{\,2\:\!n-2} \x1$, would require additional all-order insight 
into the corresponding coefficients in the crucial decomposition of the 
unfactorized partonic cross section (\ref{WHqgD}). One may hope that an
extension of Ref.~\cite{Anastasiou:2014vaa} to the complete N$^3$LO 
corrections will soon provide useful information also for the large-$x$
resummation of the quark-gluon channel.

\vspace{3mm}
\noindent{\bf Acknowledgments}

\noindent
We thank S. Marzani for his interest in these hitherto unpublished results 
which led to the writing of the present article. 
 This research has been supported by 
the European Research Council under Advanced Investigator Grant ERC-AdG-228301;
the German Research Foundation (DFG) through Sonderforschungsbereich 
Transregio 9, Computergest\"utzte Theoretische Teilchenphysik;
the UK Science \& Technology Facilities Council (STFC) under grant number 
ST/G00062X/1,
and the Research Executive Agency (REA) of the European Union under the Grant 
Agreement number PITN-GA-2010-264564 (LHCPhenoNet).
Our calculations were performed using the symbolic manipulation system FORM 
\cite{FORM}.

\end{document}